\documentclass[amsmath,amssymb,aip,jap,reprint]{revtex4-1}

\usepackage{graphicx}
\usepackage{dcolumn}
\usepackage{subfigure}
\usepackage{bm}

\begin{document}


\title{Elastic constants of the II-IV nitride semiconductors MgSiN$_{2}$, MgGeN$_{2}$ and MgSnN$_{2}$}

\author{M. R{\aa}sander}\email{m.rasander@imperial.ac.uk}
\affiliation{%
Applied Physics, Division of Materials Science, Department of Engineering Sciences and Mathematics, Lule{\aa} University of Technology, 971 87 Lule{\aa}, Sweden}
\affiliation{%
Department of Materials, Imperial College London, Exhibition Road, London, SW7 2AZ, United Kingdom
}%
\author{M. A. Moram}
\affiliation{%
Department of Physics, Cavendish Laboratory, University of Cambridge, JJ Thomson Avenue, Cambridge, CB3 0HE, United Kingdom
}%
\date{\today}

\begin{abstract}
The single crystal elastic constants, polycrystalline elastic moduli and related properties of orthorhombic MgSiN$_{2}$, MgGeN$_{2}$ and MgSnN$_{2}$ have been calculated using density functional theory and compared to the related wurtzite structured AlN, GaN and InN. Since there are no experimental studies of single crystal elastic properties of neither MgSiN$_{2}$, MgGeN$_{2}$ or MgSnN$_{2}$, we have established the accuracy of the calculations by comparison with experimental data for AlN, GaN and InN. The calculated polycrystalline elastic moduli of MgSiN$_{2}$ are found to be in good agreement with available experimental elastic moduli. It will be shown that MgSiN$_{2}$ and MgGeN$_{2}$ have a small $xy$-plane lattice mismatch with AlN and GaN, respectively, while at the same time being significantly softer than both AlN and GaN. This shows that MgSiN$_{2}$ and MgGeN$_{2}$ should be possible to be grown on AlN and GaN without significant lattice mismatch or strain.
\end{abstract}

\maketitle

\section{Introduction}
The Group-III-nitride semiconductors AlN, GaN and InN are widely used in optoelectronics and high power electronic devices.\cite{Nakamura2000,Kramers2007,Mishra2008} AlN and its alloys are also used widely in energy harvesting devices and RF applications.\cite{Elfrink2009} However, improved efficiencies are required for III-nitride-based ultraviolet (UV) light emitting diodes\cite{Kneissl2011}, solar cells\cite{Wu2009} and energy harvesting devices.\cite{Elfrink2009} Group II-IV nitride semiconductors are of growing interest in this regards, as their bonding and crystal structures are related to those of the III-nitrides but they offer different combinations of band gaps and lattice parameters, thereby opening up additional possibilities for device design.\cite{Paudel2009,Punya2011} For example: Zn-based II-IV nitrides, such as ZnSnN$_{2}$, are of current interest for solar cells,\cite{Paudel2009,Punya2011} whereas wide band gap II-IV nitrides, such as MgSiN$_{2}$ and MgGeN$_{2}$, may find applications as part of UV optoelectronic devices.\cite{Quirk,Rasander2016,Punya2016}
\par
Before new materials can be useful for applications it is required to establish the fundamental physical properties of these materials. Fundamental materials parameters are needed to design, characterise and simulate new devices effectively. For example, accurate elastic constants are essential for determining the composition of epitaxial films using X-ray difraction\cite{Moram2009} and for assessing the critical thicknesses for strain relaxation in device heterostructures.\cite{Holec2007}
\par
In this study, we will investigate the elastic properties of Mg-IV-N$_{2}$ for IV~=~Si, Ge and Sn using calculations based on density functional theory. Previous theoretical studies of Mg-IV-N$_{2}$ have focused on the electronic properties of MgSiN$_{2}$,\cite{Quirk,Rasander2016,deBoer2015,Punya2016,Huang} MgGeN$_{2}$\cite{Punya2016,Huang} and MgSnN$_{2}$.\cite{Punya2016} Recently, there has also been studies of the lattice dynamics of MgSiN$_{2}$ and MgGeN$_{2}$,\cite{Rasander2017,Pramchu2017} as well as of the thermal expansion in MgSiN$_{2}$.\cite{Rasander2017b} Both MgSiN$_{2}$ and MgGeN$_{2}$ powders have been found experimentally\cite{Bruls1,David1970,Quirk} and shown to have an orthorhombic crystal structure which is derived from the wurtzite structure. In the case of MgSiN$_{2}$, polycrystalline elastic constants\cite{Bruls2} as well as derived properties,\cite{Bruls2001} such as sound velocities and the Debye temperature, have been measured. Arab {\it et al.}\cite{Arab2016} have calculated the elastic constants of MgSiN$_{2}$ previously, however, in their study Arab {\it et al.} were more interested in how the elastic properties varied under pressure than comparing the elastic properties of MgSiN$_{2}$ with other compounds. Here we will present the elastic properties of MgSiN$_{2}$ as well as MgGeN$_{2}$ and MgSnN$_{2}$. In addition to presenting the elastic constants of the Mg-IV-N$_{2}$ systems, we will critically compare the elastic properties of these systems with wurtzite AlN, GaN and InN. In addition, we will also make comparisons to the Zn-based II-IV nitrides ZnSiN$_{2}$, ZnGeN$_{2}$ and ZnSnN$_{2}$ for which calculations of the elastic properties have been performed by Paudel and Lambrecht.\cite{Paudel2009}

\section{Methods}
\par
Density functional calculations have been performed using the projector augmented wave (PAW) method\cite{Blochl} as implemented in the Vienna {\it ab initio} simulations package (VASP).\cite{KresseandFurth,KresseandJoubert} We have used the PBEsol generalized gradient approximation\cite{PBEsol} for the exchange-correlation energy functional. The plane wave energy cut-off was set to 800~eV and we have used $\Gamma$-centered k-point meshes with the smallest allowed spacing between k-points of 0.1~\AA$^{-1}$. The atomic positions and simulation cell shapes were relaxed until the Hellmann-Feynman forces acting on atoms were smaller than 0.001~eV/\AA.
\par
\begin{figure*}[t]
\includegraphics[width=5.5cm]{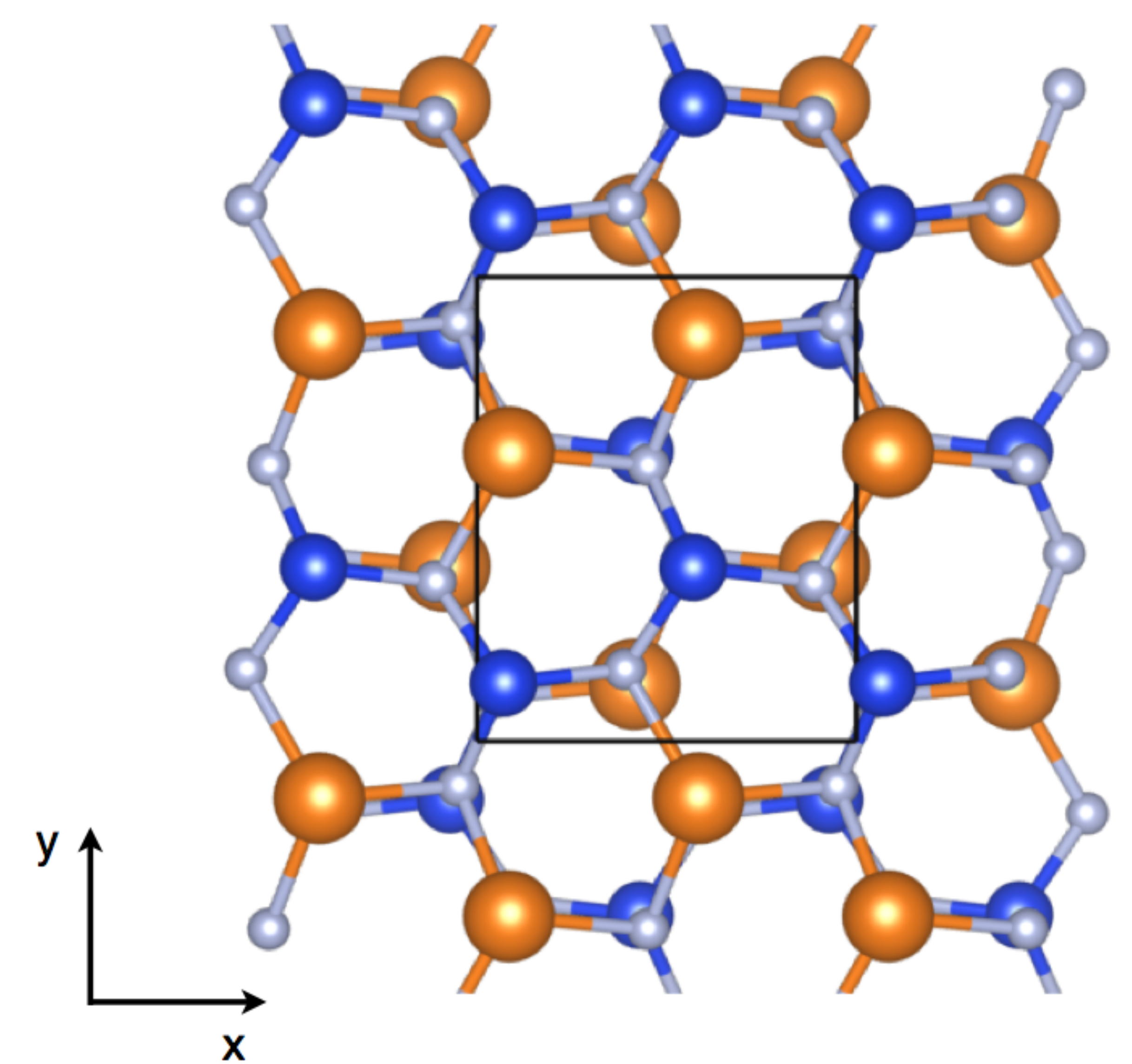}\includegraphics[width=6.5cm]{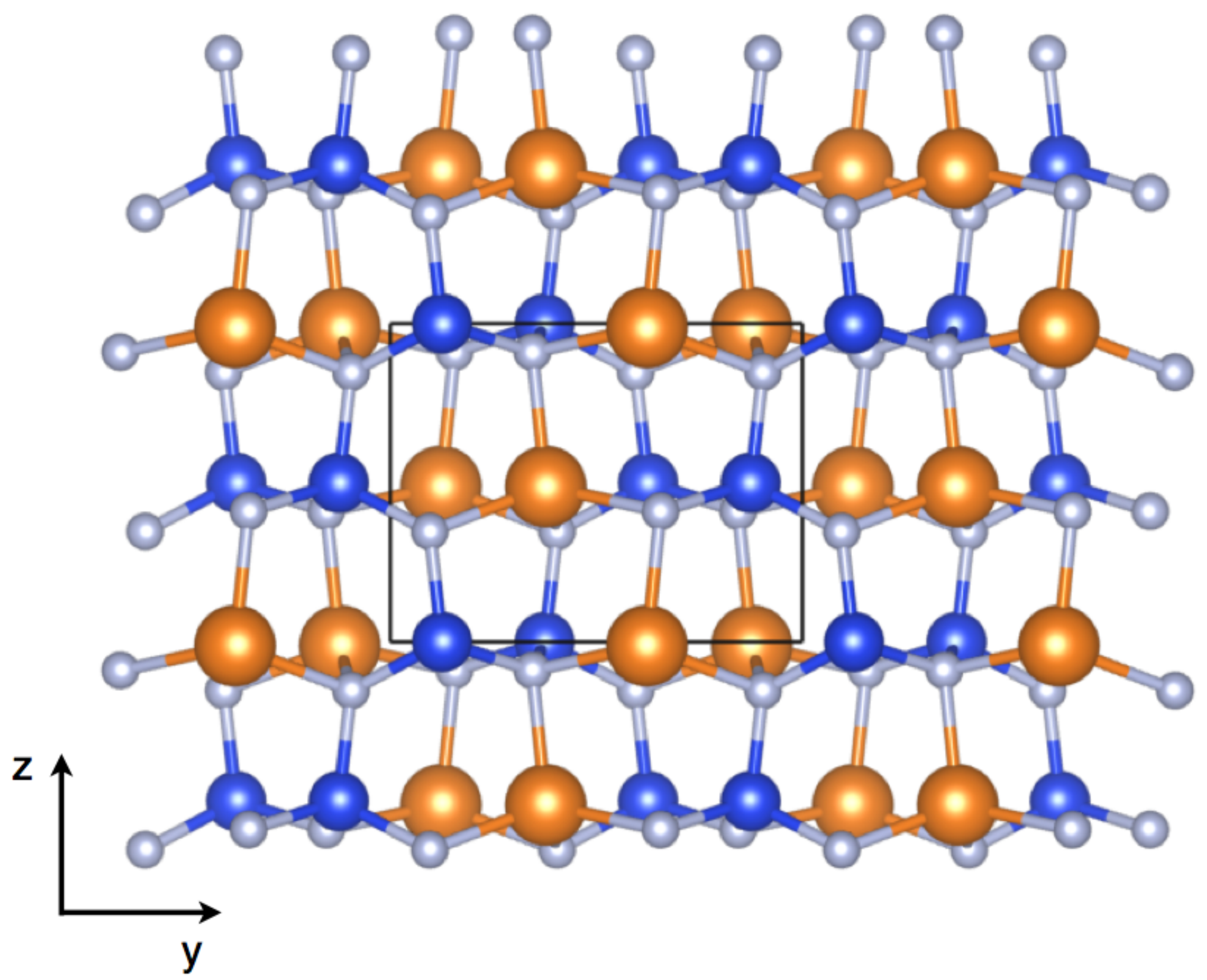}\includegraphics[width=5.5cm]{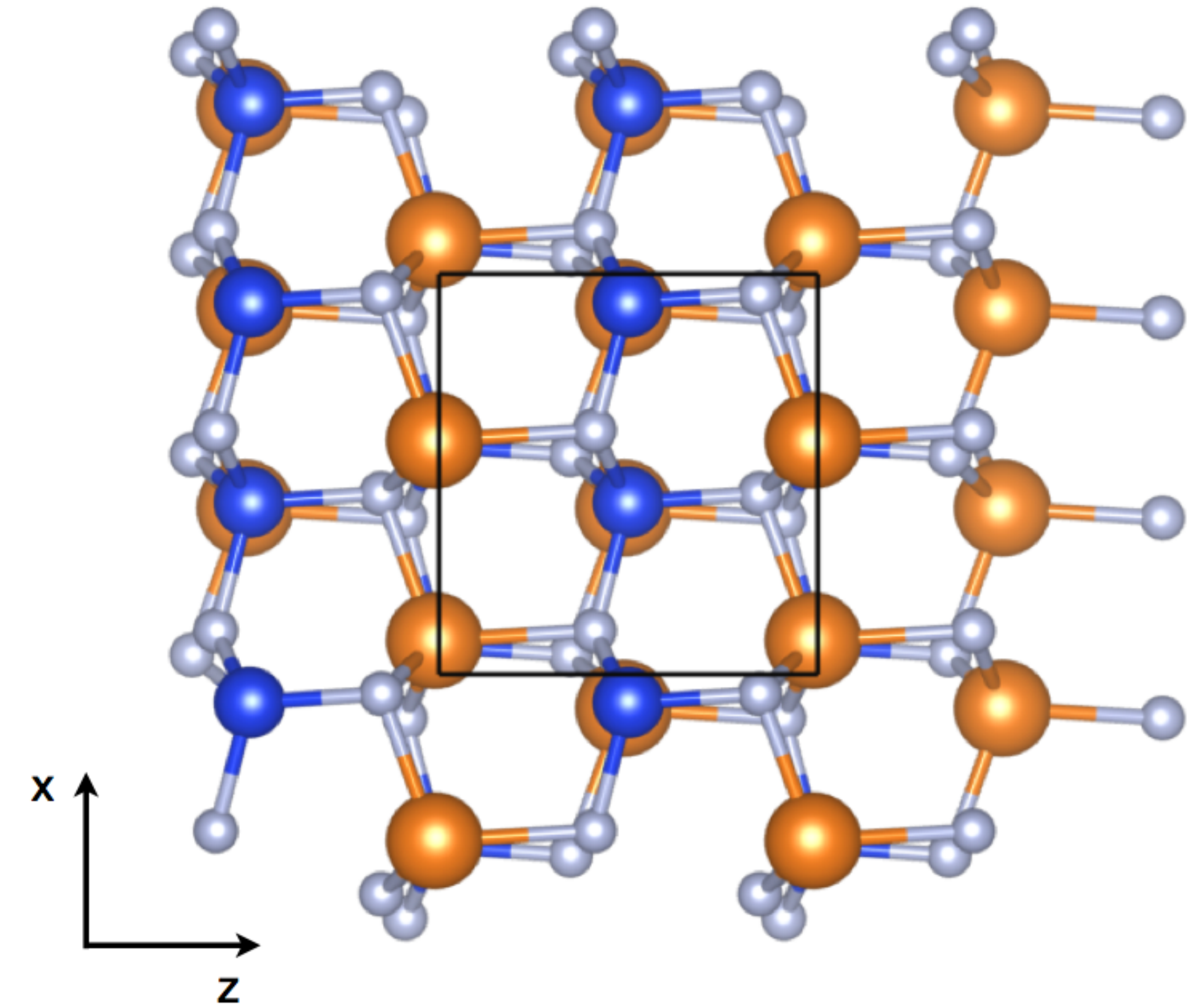}
\caption{\label{fig:planes} (Color online) The crystal structure of Mg-IV-N$_{2}$ shown along the x-y, y-z and x-z planes. Mg, IV and N are shown in bronze, blue and grey spheres respectively. The solid black rectangles show the boundaries of the primitive unit cell.}
\end{figure*}
\begin{table}[b]
\caption{\label{tab:structure-MgSiN2} Lattice constants and crystallographic coordinates $(x,y,z)$ of orthorhombic MgSiN$_{2}$.}
\begin{ruledtabular}
\begin{tabular}{lccc}
XC & $a$ (\AA) & $b$ (\AA) & $c$ (\AA) \\
 \hline
PBEsol & 5.280 & 6.480 & 4.999\\
 Expt. ($T=10$~K)\cite{Bruls1} & 5.27078(5) & 6.46916(7) & 4.98401(5) \\
 \hline
   \multicolumn{4}{c}{PBEsol} \\
  & $x$ & $y$ & $z$ \\
 \hline
 Mg & 0.0840 &  0.6227 &  0.9883\\
 Si & 0.0701 & 0.1255 &  0.0000\\
 N(1) & 0.0481 & 0.0955 & 0.3473\\
 N(2) & 0.1090 & 0.6556 & 0.4075\\
 \hline
 \multicolumn{4}{c}{Expt. ($T=10$~K)\cite{Bruls1}}\\
& $x$ & $y$ & $z$ \\
\hline
Mg & 0.08448(34) & 0.62255(30) & 0.9866(5) \\
Si & 0.0693(5) & 0.1249(4) & 0.0000\\
N(1) & 0.04855(17) & 0.09562(15) & 0.3472(4)\\
N(2) & 0.10859(18) & 0.65527(14) & 0.4102(4)\\
 \end{tabular}
\end{ruledtabular}
\end{table}
\begin{table}[b]
\caption{\label{tab:structure-MgGeN2} Lattice constants and crystallographic coordinates $(x,y,z)$ of orthorhombic MgGeN$_{2}$. Here we have shifted the $z$ values such that the $z$ coordinate of Ge is 0.}
\begin{ruledtabular}
\begin{tabular}{lccc}
XC & $a$ (\AA) & $b$ (\AA) & $c$ (\AA) \\
 \hline
PBEsol & 5.497 & 6.618 & 5.172 \\
 Expt.\cite{David1970} & 5.494 & 6.611 & 5.166 \\
 \hline
   \multicolumn{4}{c}{PBEsol} \\
  & $x$ & $y$ & $z$ \\
 \hline
 Mg & 0.0845 & 0.6233  & 0.9913 \\
 Ge & 0.0736 & 0.1258 & 0.9988 \\
 N(1) & 0.0596 & 0.1067 & 0.3588\\
 N(2) & 0.1002 & 0.6438 & 0.3971\\
  \hline
   \multicolumn{4}{c}{Expt.\cite{David1970}} \\
  & $x$ & $y$ & $z$ \\
 \hline
 Mg & 0.083  & 0.625  & 0.000 \\
 Ge & 0.083 & 0.125 & 0.000 \\
 N(1) & 0.083 & 0.125  & 0.380\\
 N(2) & 0.083 & 0.625 & 0.400 \\
 \end{tabular}
\end{ruledtabular}
\end{table}
\begin{table}[b]
\caption{\label{tab:structure-MgSnN2} Lattice constants and crystallographic coordinates $(x,y,z)$ of orthorhombic MgSnN$_{2}$. Here we have shifted the $z$ values such that the $z$ coordinate of Sn is 0.}
\begin{ruledtabular}
\begin{tabular}{lccc}
XC & $a$ (\AA) & $b$ (\AA) & $c$ (\AA) \\
 \hline
PBEsol & 5.907 & 6.878 & 5.472\\
\hline
   \multicolumn{4}{c}{PBEsol} \\
  & $x$ & $y$ & $z$ \\
 \hline
 Mg & 0.0832 & 0.6245  & 0.9942 \\
 Sn & 0.0832 & 0.1257 & 0.9970 \\
 N(1) & 0.0786 & 0.1222 & 0.3764\\
 N(2) & 0.0851 & 0.6274 & 0.3785\\
 \end{tabular}
\end{ruledtabular}
\end{table}
\par
The elastic constants have been evaluated following Refs.~\onlinecite{LePage} and \onlinecite{Wu} where the stress is evaluated from the application of a strain to the system and the elastic constants are evaluated from Hook's law ${\bm \sigma}=\bar{\bm C}{\bm \epsilon}$, where ${\bm \sigma}$ is the stress, ${\bm \epsilon}$ is the small applied strain and $\bar{\bm C}$ is the elastic constants tensor. This approach has been used successfully to calculate the elastic constants of both zinc-blende\cite{n} and wurtzite\cite{o,p} Group III-nitride alloys, producing accurate elastic constants with a relatively low computational cost.\cite{LePage,o,r,Rasander2015} Here we evaluate the elastic constants for orthorhombic Mg-IV-N$_{2}$, where IV = Si, Ge and Sn, (space group Pna2$_{1}$) illustrated in Fig.~\ref{fig:planes} and wurtzite structured AlN, GaN and InN. For an orthorhombic crystal there are 9 independent elastic constants which in the Voigt notation, where $1=xx$, $2=yy$, $3=zz$, $4=yz$, $5=zx$ and $6=xy$, are given by $c_{11}$, $c_{22}$,  $c_{33}$, $c_{12}$, $c_{13}$, $c_{23}$, $c_{44}$, $c_{55}$ and $c_{66}$. The wurtzite structure has 5 independent elastic constants $c_{11}$, $c_{33}$, $c_{12}$, $c_{13}$ and $c_{44}$, with a sixth elastic constant defined as $c_{66} = (1/2)(c_{11}-c_{12})$.

\section{Structural properties}
The orthorhombic crystal structure shown in Fig.~\ref{fig:planes} can be derived from the wurtzite structure by substituting one Group II and one Group IV atom for every two Group III atoms. The actual transformation follows as ${\bm a}={\bm a}_{1}+2{\bm a}_{2}$ and ${\bm b}=2{\bm a}_{1}$, where ${\bm a}_{1}$ and ${\bm a}_{2}$ are the lattice constants of the wurtzite lattice in the $xy$-plane while ${\bm a}$ and ${\bm b}$ are the lattice vectors in the $xy$-plane of the orthorhombic structure. The ${\bm c}$ lattice vector is common to both crystal structures. In the orthorhombic Pna2$_{1}$ structure all atoms occupy the $4a$ wyckoff crystal positions, with crystal coordinates $x$, $y$ and $z$. In Tables~\ref{tab:structure-MgSiN2}, \ref{tab:structure-MgGeN2} and \ref{tab:structure-MgSnN2}, we show the lattice constants and obtained wyckoff positions for the orthorhombic Mg-IV-N$_{2}$ systems. In Table~\ref{tab:wurtzite-structure} we show the calculated lattice constants of wurtzite AlN, GaN and InN, together with experimental reference values. Overall, we find that the calculated structural parameters are in very good agreement with reported experimental structures for both Group II-IV and Group III nitrides, with differences of the order of less than 1\% which is to be expected when using the PBEsol density functional.\cite{PBEsol,Rasander2015} We also note that our calculated lattice constants for the II-IV nitrides are in reasonable agreement with previous work in which the electronic structure of MgSiN$_{2}$ was calculated using local and semi-local approximations for the exchange-correlation energy functional.\cite{Punya2016,Fang1,Huang}
\par
We note that the atoms in the orthorhombic Mg-IV-N$_{2}$ structure form a slightly distorted tetrahedral framework compared to pure wurtzite crystal structures which is clearly shown in Fig.~\ref{fig:planes} and discussed in detail in Refs. \onlinecite{Rasander2017} and \onlinecite{Bruls1}. This distortion of the atomic arrangement is accomplished in order to accommodate the size difference as well as interatomic bond strengths between the Group II and IV elements in the II-IV nitride systems. 
\par
Since the orthorhombic structures are derived from the wurtzite structure, it is possible to define a wurtzite-like lattice constant for the orthorhombic structures, $\bar{a}_{w} = (a/\sqrt{3}+b/2)/2$,\cite{Rasander2017} as well as the deviation from a hexagonal lattice, $\Delta_{w} = |a\sqrt{3}-b/2|/\bar{a}_{w}$.\cite{Rasander2017} Based on our calculations we find the average wurtzite-like lattice constant to be 3.144~\AA, 3.241~\AA~and 3.425~\AA~for MgSiN$_{2}$, MgGeN$_{2}$ and MgSnN$_{2}$, respectively. The deviation from hexagonality is found to be  6.1\%, 4.2\% and 0.8\% for MgSiN$_{2}$, MgGeN$_{2}$ and MgSnN$_{2}$, respectively. MgSiN$_{2}$ therefore shows the largest deviation from a hexagonal structure while MgSnN$_{2}$ shows only a very small deviation. Compared to the III-nitrides, we note that $\bar{a}_{w}$ in MgSiN$_{2}$ is slightly larger than the in-plane lattice constant in AlN (3.144~\AA~vs. 3.112~\AA), while $\bar{a}_{w}$ in MgGeN$_{2}$ is slightly larger than the in-plane lattice constant in GaN (3.241~\AA~vs. 3.179~\AA). In MgSnN$_{2}$ $\bar{a}_{w}$ is intermediate between the in-plane lattice constant in GaN and InN (3.425~\AA~vs. 3.179~\AA~and 3.531~\AA, respectively). We also note that the wurtzite-like lattice constant in MgSiN$_{2}$ is only slightly smaller than the in-plane lattice constant in GaN (3.144~\AA~vs. 3.179~\AA), and, in fact, the wurtzite-like lattice constant in MgSiN$_{2}$ is found halfway between the in-plane lattice constant of AlN and GaN.
 
\par
\begin{table}[t]
\caption{\label{tab:wurtzite-structure} Calculated crystal structure parameters for the Group-III nitrides AlN, GaN and InN.}
\begin{ruledtabular}
\begin{tabular}{lccc}
 & $a$ (\AA)& $c$ (\AA) & $c/a$ \\
 \hline
AlN (PBEsol) & 3.112 & 4.980 & 1.600\\
AlN (Expt.\cite{Wu2009}) & 3.112 & 4.982 & 1.601\\
GaN (PBEsol) & 3.179 & 5.181 & 1.630\\
GaN (Expt.\cite{Wu2009}) & 3.189 & 5.185 & 1.626\\
GaN (Expt.\cite{Leszczynski1996}) & 3.189 & 5.1864 & 1.626\\
InN (PBEsol) & 3.531 & 5.704 & 1.615\\
InN (Expt.\cite{Wu2009}) & 3.533 & 5.693 & 1.611\\
InN (Expt.\cite{Osamura1975}) & 3.544 & 5.718 & 1.613 \\
 \end{tabular}
\end{ruledtabular}
\end{table}
\begin{table*}[t]
\caption{\label{tab:elasticconstants} Single crystal elastic constants of MgSiN$_{2}$, MgGeN$_{2}$, MgSnN$_{2}$, AlN, GaN and InN. All elastic constants are given in GPa. For AlN, GaN and InN experimental elastic constants are also shown.}
\begin{ruledtabular}
\begin{tabular}{ccccccccccc}
& \multicolumn{1}{c}{MgSiN$_{2}$} & \multicolumn{2}{c}{AlN} & \multicolumn{1}{c}{MgGeN$_{2}$} & \multicolumn{2}{c}{GaN} & \multicolumn{1}{c}{MgSnN$_{2}$} & \multicolumn{3}{c}{InN}\\
\cline{3-4}\cline{6-7}\cline{9-11}
$c_{ij}$ & PBEsol  & PBEsol &  Expt.\cite{McNeil} & PBEsol & PBEsol & Expt.\cite{Polian} & PBEsol & PBEsol & Expt. \cite{Sheleg} & Expt.\cite{Morales2009}\\
\hline
$c_{11}$ & 310.2  & 382.4  & 411$\pm10$ & 268.5 & 346.5 & 390 & 223.5 & 216.9 & 190(7) & 237(7)\\
$c_{22}$ &  303.6 & $=c_{11}$  &  $=c_{11}$ & 264.5 & $=c_{11}$ & $=c_{11}$ & 210.7 & $=c_{11}$ & $=c_{11}$ & $=c_{11}$\\
$c_{33}$ &  320.2 &  356.7 & 389$\pm10$ & 274.6 & 385.2 & 398 & 210.2 & 224.9 & 182(6) & 236(6)\\
$c_{12}$ &  148.8  &  138.8 & 149$\pm10$ & 126.0 & 127.8 &  145 & 98.9 & 107.4 & 104(3) & 106(4)\\
$c_{13}$ &  75.8  &  108.7  & 99$\pm4$ & 76.7 & 91.8 & 106 & 79.0 & 88.2 & 121(7) & 85(3)\\
$c_{23}$ &  116.8 &  $=c_{13}$  & $=c_{13}$ & 99.8 & $=c_{13}$ & $=c_{13}$ & 80.8 & $=c_{13}$ & $=c_{13}$ & $=c_{13}$\\
$c_{44}$ &  121.5 & 112.0 &  125$\pm5$ & 92.8 & 92.3 & 105 & 61.0 & 48.9 & 10(1) & 53(3) \\
$c_{55}$ &  84.2 & $=c_{44}$  &  $=c_{44}$ & 71.2 & $=c_{44}$ & $=c_{44}$ & 56.2 & $=c_{44}$ & $=c_{44}$ & $=c_{44}$\\
$c_{66}$ &  128.6 &  121.8  & 131$\pm 10$ & 99.4 & 109.4 & 123 & 64.2 & 54.7 & 43 & 66 \\
 \end{tabular}
\end{ruledtabular}
\end{table*}
\begin{table}[t]
\caption{\label{tab:zn-iv-n2} Single crystal elastic constants of ZnSiN$_{2}$, ZnGeN$_{2}$ and ZnSnN$_{2}$. All elastic constants are given in GPa. Data taken from reference \onlinecite{Paudel2009}.}
\begin{ruledtabular}
\begin{tabular}{lccc}
$c_{ij}$ & ZnSiN$_{2}$ & ZnGeN$_{2}$ &  ZnSnN$_{2}$\\
 \hline
$c_{11}$ & 408 & 358 & 290\\
$c_{22}$ & 383 & 341 & 272\\
$c_{33}$ & 463 & 401 & 306\\
$c_{12}$ & 146 & 136 & 128\\
$c_{13}$ & 105 & 98 & 105\\
$c_{23}$ & 117 & 103 & 100\\
$c_{44}$ &110 & 95 & 67\\
$c_{55}$ &104 & 86 & 64\\
$c_{66}$ & 124 & 105 & 74\\
 \end{tabular}
\end{ruledtabular}
\end{table}

\section{Single crystal elastic constants}

\par
Table~\ref{tab:elasticconstants} shows the elastic constants for MgSiN$_{2}$, MgGeN$_{2}$ and MgSnN$_{2}$, as well as for AlN, GaN and InN. The PBEsol approximation has previously been found to obtain very accurate elastic constants,\cite{Rasander2015} and as can be seen for AlN, GaN and InN in Table~\ref{tab:elasticconstants}, we find that the present PBEsol calculations are in very good agreement with available experimental data. The general trend is that the calculations underestimate the elastic constants for the Group III-nitrides slightly. We also reproduce the trend that the elastic constants become smaller as the Group III element is varied from Al to Ga to In. Note that in the case of InN, we show different sets of experimental reference values, which differs from each other. In the older study by Sheleg and Savastenko,\cite{Sheleg} the elastic constants in InN are consistently smaller than in the study by Morales {\it et al.}\cite{Morales2009} Especially the $c_{44}$ elastic constant is very small according to Sheleg and Savastenko.\cite{Sheleg} The calculated PBEsol results are in very good agreement with the elastic constants measured by Morales {\it et al.}\cite{Morales2009} 
\par
For the Mg-IV-N$_{2}$ systems, we find that the elastic constants become smaller as the Group IV element is varied from Si to Sn, i.e. the same trend as for the Group III nitrides. However, the reduction in the elastic constants are less dramatic. We also note that MgSiN$_{2}$ and MgGeN$_{2}$ are softer than AlN and GaN, respectively, where the elastic constants of the Mg-IV-N$_{2}$ systems are generally softer or even significantly softer than in the corresponding Group III nitride. In fact, the $c_{11}$, $c_{22}$ and $c_{33}$ elastic constants in MgSiN$_{2}$ are even smaller than the $c_{11}$ and $c_{33}$ elastic constants in GaN. We also find that the elastic constants in MgSnN$_{2}$ are similar in size to the elastic constants in InN. That the elastic constants in MgSiN$_{2}$ and MgGeN$_{2}$ are softer than in the corresponding Group III nitrides is beneficial since it will make it possible to grow these nitride phases on substrates with slightly expanded or contracted lattice constants with smaller strains than what is possible with the Group III nitrides. This is interesting since reducing strain in thin film growth will assist in reducing the presence of dislocations in the films.
\par
There have been previous investigations of the elastic constants of the similar Zn-IV-N$_{2}$ systems and it is interesting to see how the elastic constants of these systems compare to the elastic constants of the Mg-IV-N$_{2}$ systems. In Table~\ref{tab:zn-iv-n2} we show the elastic constants of ZnSiN$_{2}$, ZnGeN$_{2}$ and ZnSnN$_{2}$ calculated by Paudel and Lambrecht\cite{Paudel2009} using the local density approximation (LDA). We note that the LDA has a tendency of overestimating atomic bond strengths which results in too high elastic constants compared to experiments.\cite{Rasander2015,Csonka2009}  Even so, it appears that the Mg-IV-N$_{2}$ systems are significantly softer than the corresponding Zn-IV-N$_{2}$ system. 
\par

\begin{table*}[t]
\caption{\label{tab:elasticmoduli} Calculated shear modulus and bulk modulus evaluated using both Voigt and Reuss approximations, Hill's average shear and bulk modulus, Young's modulus and Poisson's ratio as defined in the text, Eqs.~(\ref{eq:Voigtshear}) to (\ref{eq:Reussbulk}). The experimental references for which both Reuss and Voight approximations are found are evaluated from the single crystal elastic constants given in Table~\ref{tab:elasticconstants}. All moduli are given in GPa. }
\begin{ruledtabular}
\begin{tabular}{lccccccccc}
 & $G_{R}$ & $G_{V}$ & $B_{R}$ & $B_{V}$ & $G$ & $B$ & $B/G$ & $E$ & $\nu$ \\
\hline
 \multicolumn{10}{c}{MgSiN$_{2}$} \\
 \cline{2-10}
PBEsol & 101.2 & 106.4 & 178.6 & 179.6 & 103.8 & 179.1 & 1.726 & 261.0 & 0.257\\
Expt.\cite{Bruls2} & -& -& -& -&  113 & 184 & 1.63 & 281 & 0.246 \\
\hline
 \multicolumn{10}{c}{MgGeN$_{2}$} \\
 \cline{2-10}
PBEsol & 84.0 & 86.4 & 156.5 & 157.0 & 85.2 & 156.7 & 1.840 & 216.3 &  0.270 \\
\hline
 \multicolumn{10}{c}{MgSnN$_{2}$} \\
 \cline{2-10}
PBEsol & 61.6 & 62.0 & 128.7 & 129.1 &  61.8 & 128.9 & 2.085 & 159.9 & 0.293\\
\hline
 \multicolumn{10}{c}{AlN} \\
  \cline{2-10}
PBEsol & 119.6 & 120.2 & 202.9 & 203.8 & 119.9 & 203.3 & 1.696 & 300.6 & 0.254\\
Expt.\cite{McNeil} & 133 & 134 & 210 & 212 & 133 & 211 & 1.58 & 330 & 0.239 \\
Expt.\cite{Tsubouchi} & -&- &- &- & 117 & 202 & 1.73 & 295 & 0.257\\
Expt.\cite{Boch} & - & -& -&- & 126 & 206 & 1.63 & 315 & 0.246\\
Expt.\cite{Gerlich} & - & -& -&- & 131 & 160 & 1.22 & 308 & 0.178\\
\hline
 \multicolumn{10}{c}{GaN} \\
  \cline{2-10}
PBEsol & 106.9 & 109.9 & 190.0 & 190.0 & 108.4 & 190.0 & 1.743 & 273.0 & 0.259\\
Expt.\cite{Polian} & 119 & 121 & 210 & 210 & 120 & 210 & 1.75 & 303 & 0.260\\
\hline
 \multicolumn{10}{c}{InN} \\
  \cline{2-10}
PBEsol & 54.5 & 55.5 & 136.2 & 136.3 & 55.0 & 136.2 & 2.476 & 145.5 & 0.322\\
Expt.\cite{Sheleg} & 18 & 27 & 139 & 139 & 22 & 139 & 6.22 & 64 & 0.424\\
Expt.\cite{Morales2009} & 62 & 63 & 140 & 140 & 63 & 140 & 2.238 & 163 & 0.306\\
 \end{tabular}
\end{ruledtabular}
\end{table*}

\section{Polycrystalline elastic constants}
So far, all experimental studies of MgSiN$_{2}$ and MgGeN$_{2}$ have been performed using powders and therefore there are no single crystal studies to compare our calculated elastic constants for the Mg-IV-N$_{2}$ systems with. Although the single crystal elastic constants, $c_{ij}$, cannot be determined, some bulk elastic properties have been determined experimentally for MgSiN$_{2}$ and are available for comparison to calculations. Here, the polycrystalline bulk moduli, $B$, and shear moduli, $G$ have been determined from the elastic constants calculated for the Mg-IV-N$_{2}$ and the III-N systems using the Voigt-Reuss-Hill scheme\cite{Grimvall,Voigt,Reuss,Hill} as follows: For a general crystal symmetry the Voigt, $G_{V}$, and Reuss, $G_{R}$, shear moduli are given by
\begin{multline}\label{eq:Voigtshear}
G_{V}= \frac{1}{15}\left(c_{11} + c_{22} + c_{33} -c_{12} - c_{13} - c_{23}  \right)  \\
 +\frac{1}{5}\left(c_{44} + c_{55} + c_{66}\right),
\end{multline}
and
\begin{multline}\label{eq:Reussshear}
G_{R}= 15 \Bigl\{ 4(s_{11}+ s_{22} + s_{33}) \\ 
- 4(s_{12} + s_{13}+ s_{23})  + 3(s_{44} + s_{55} + s_{66})\Bigr\}^{-1}
\end{multline}
respectively. The $s_{ij}$'s are the elastic compliance constants, which are evaluated as the inverse of the elastic constants tensor, $\bar{S} = \bar{C}^{-1}$. The Voigt, $B_{V}$, and Reuss, $B_{R}$, bulk modulus are given by
\begin{equation}\label{eq:Voigtsbulk}
B_{V} = \frac{1}{9}\left(c_{11} + c_{22} + c_{33} \right) + \frac{2}{9}\left(c_{12} + c_{13} + c_{23} \right),
\end{equation}
and
\begin{equation}\label{eq:Reussbulk}
B_{R} = \frac{1}{\left(s_{11} + s_{22} + s_{33} \right) + 2\left(s_{12} + s_{13} + s_{23} \right)},
\end{equation}
respectively.\cite{Grimvall} Since the Voigt and Reuss approximations represent the upper and lower bounds to the polycrystalline elastic bulk and shear modulus, Hill\cite{Hill} proposed to use the arithmetic mean of the Voigt and Reuss extremes, i.e. $G=(G_{V}+G_{R})/2$ and $B = (B_{V} + B_{R})/2$. In terms of these elastic moduli the Young's modulus, $E$, and Poisson's ratio, $\nu$, may be determined from standard relations between the Young's modulus, Poisson's ratio, bulk modulus and shear modulus.\cite{Grimvall} 
\par
The evaluated polycrystalline elastic moduli for MgSiN$_{2}$, MgGeN$_{2}$ and MgSnN$_{2}$ are compared to AlN, GaN and InN in Table~\ref{tab:elasticmoduli}. It is clear that both the bulk and shear modulus decrease as the crystals become heavier for both Mg-IV-N$_{2}$ and III-N systems. The same is also true for the Young's modulus. Furthermore, we note that when comparing Mg-IV-N$_{2}$ and III-N systems the bulk modulus, shear modulus and Young's modulus of MgSiN$_{2}$, MgGeN$_{2}$ and MgSnN$_{2}$ are smaller than in both AlN and GaN. MgSnN$_{2}$ also has a smaller bulk modulus than InN, while both the shear modulus and Young's modulus in MgSnN$_{2}$ are larger than in InN. 
\par
In Table~\ref{tab:elasticmoduli}, we also present measured elastic moduli of powder samples of MgSiN$_{2}$\cite{Bruls2} and AlN.\cite{Tsubouchi,Boch,Gerlich} We find that our calculations are generally in very good agreement with these experimental data. The calculated shear, bulk and Young's modulus are all slightly smaller than the experimental observations which is in line with the previously discussed small underestimation of the single crystal elastic constants.
\par
We note that the shear modulus represents the resistance towards plastic deformation while the bulk modulus measures the resistance to fracture. A large shear modulus is also significative of a higher degree of directional bonding between atoms in the crystal. It is therefore clear, since the largest shear modulus and bulk modulus are obtained for MgSiN$_{2}$ and AlN among the two forms of nitrides, that the resistance towards both plastic deformation and fracture become lower as the nitride systems become heavier. The same trend is also valid for the stiffness of the Mg-IV-N$_{2}$ and III-N systems, since the Young's modulus is the largest within each nitride group for MgSiN$_{2}$ and AlN, respectively.
\par
In Table~\ref{tab:elasticmoduli} we also present the quotient $B/G$, which was introduced by Pugh.\cite{Pugh} A high (low) value for $B/G$ is related to a ductile (brittle) behaviour. The critical value which separates ductile from brittle behaviour is 1.75.\cite{Pugh} As can be seen in Table~\ref{tab:elasticmoduli}, MgSiN$_{2}$, AlN and GaN all fall below $B/G$ of 1.75 and are therefore brittle; GaN is very much on the borderline, while MgSiN$_{2}$ and AlN are equally brittle. The remaining Mg-IV-N$_{2}$ systems and InN are all above 1.75 and are therefore considered to be more ductile. We note that the $B/G$ value for InN obtained by using the single crystal elastic constants obtained by Sheleg and Savastenko\cite{Sheleg} is very large which is due to the significantly underestimated shear modulus compared to both our calculations and the experiment by Morales {\it et al.}\cite{Morales2009}
\par
The Poisson's ratio, $\nu$, measures the stability of a crystal against shear and is also associated with the volume change during uniaxial deformation. $\nu=0.25$ and $\nu=0.5$ are the lower and upper limits for central-force solids. Furthermore, for $\nu=0.5$ no volume change occurs during elastic deformation and this limit also represent the case of infinite elastic anisotropy. A low value for the Poisson's ratio ($\nu<0.25$) suggests that the interatomic forces in the solid are non-central. We find that both MgSiN$_{2}$ and AlN have Poisson's ratios that are close to 0.25 and therefore close to the lower limit of central-force solids. As the Group IV and Group III element are varied the Poisson's ratio increases and reaches a maximum of 0.293 and 0.322 for MgSnN$_{2}$ and InN, respectively.

\section{Elastic anisotropy}
\begin{table}[t]
\caption{\label{tab:anisotropy} Calculated elastic anisotropies $A_{B}$, $A_{G}$ and $A_{L}$. The experimental references are evaluated using the experimental elastic constants presented in Table~\ref{tab:elasticconstants}.}
\begin{ruledtabular}
\begin{tabular}{lccc}
 & $A_{B}\times100$~\% & $A_{G}\times100$~\% & $A_{L}$ \\
 \hline
MgSiN$_{2}$ (PBEsol) & 0.3\% & 2.5\% & 2.35 \\
MgGeN$_{2}$ (PBEsol) & 0.2\% & 1.4\% &  2.30\\
MgSnN$_{2}$ (PBEsol) & 0.2\% & 0.3\% & 2.25 \\
AlN (PBEsol)& 0.2\% & 0.2\% & 2.25 \\
AlN (Expt.\cite{McNeil}) & 0.3\% & 0.4\% & 2.25 \\
GaN (PBEsol)& 0.0\% & 1.4\% & 2.30\\
GaN (Expt.\cite{Polian}) & 0.1\% & 0.9\% & 2.28 \\
InN (PBEsol)& 0.0\% & 0.9\% & 2.28 \\
InN (Expt.\cite{Sheleg}) & 0.0\% & 20.6\% & 3.39\\
InN (Expt.\cite{Morales2009}) & 0.1\% & 1.2\% & 2.29\\
 \end{tabular}
\end{ruledtabular}
\end{table}

Apart from knowing the size of the elastic constants, it is also important to have an understanding of the elastic anisotropy in new compounds since anisotropy is an origin for microcracks or other types of defects.\cite{Tvergaard1988} For an orthorhombic crystal, the criterion for an elastically isotropic material is that there are only two independent elastic constants which are related by $c_{11}=c_{22}=c_{33}$, $c_{12}=c_{13}=c_{23}$ and $c_{44}=c_{55}=c_{66}$ with the additional requirement that $c_{11}-c_{12}=2c_{44}$. This is clearly not fulfilled for the Mg-IV-N$_{2}$, especially for the $c_{12}$, $c_{13}$ and $c_{23}$ as well as among the $c_{44}$, $c_{55}$ and $c_{66}$ elastic constants. For a hexagonal crystal the isotropic condition is fulfilled if $c_{11}=c_{33}$ and $c_{12}=c_{13}$ with the condition that $c_{11}-c_{12}=2c_{44}$. This is not fulfilled for any of the III-nitrides either, as shown in Table~\ref{tab:elasticconstants}.
\par
In order to quantify the elastic anisotropy further it is required to have measures of the anisotropy that are valid for any crystal symmetry. Especially if we want to make a direct comparison between the Mg-IV-N$_{2}$ and III-N systems. Chung and Buessem\cite{Chung1967} noted that a crystal is elastically isotropic when the Voigt and Reuss averages of the shear modulus are identical. It is therefore possible to define the percentage anisotropy in the bulk modulus and the shear modulus for any type of crystal structure as
\begin{equation}
A_{B} = \frac{B_{V}-B_{R}}{B_{V}+B_{R}},
\end{equation}
and 
\begin{equation}
A_{G} = \frac{G_{V}-G_{R}}{G_{V}+G_{R}},
\end{equation}
respectively. In these expressions a value of zero represents elastic isotropy and a value of 1 is the largest possible anisotropy.
\par
Recently Kube\cite{Kube2016} proposed an anisotropy index, $A_{L}$ that is valid for all crystal symmetries with the additional advantage that it provides a single measure of the elastic anisotropy of a crystal. The anisotropy index is defined in terms of the Voigt and Reuss averages of the bulk and shear modulus as
\begin{equation}
A_{L} = \sqrt{\left[ {\rm ln}\left(\frac{B_{V}}{B_{R}} \right)   \right]^2 + 5\left[ {\rm ln}\left(\frac{G_{V}}{G_{R}} \right)   \right]^2}.
\end{equation}
Note that the anisotropy index is zero if the crystal is elastically isotropic and that the more anisotropic the crystal is the larger the anisotropy index will be.\cite{Kube2016} 
\par
In Table~\ref{tab:anisotropy} we show the calculated anisotropy measures presented above. Among the Mg-IV-N$_{2}$ systems, we find that the anisotropy is the largest for MgSiN$_{2}$ and the lowest for MgSnN$_{2}$. For the III-nitrides, the anisotropy is the lowest for AlN, while the largest anisotropy is found for GaN. We note that overall the elastic anisotropy is rather small in all investigated systems, even though the anisotropy is  generally larger in the Mg-IV-N$_{2}$ systems, which is related to their more anisotropic crystal structures compared to the III-nitrides. 
\par
Compared to the Zn-IV-N$_{2}$ systems, we note that MgSiN$_{2}$ is more anisotropic than ZnSiN$_{2}$ ($A_{L}=2.29$), MgGeN$_{2}$ is less anisotropic than ZnGeN$_{2}$ ($A_{L}=3.00$) and MgSnN$_{2}$ is less anisotropic than ZnSnN$_{2}$ ($A_{L}=2.30$), even though the differences are small. Here the $A_{L}$ values for the Zn-IV-N$_{2}$ systems have been evaluated using the elastic constants shown in Table~\ref{tab:zn-iv-n2}.

\section{Speed of sound and the Debye temperature}
\begin{table*}[t]
\caption{\label{tab:debye} Calculated densities, speed of sound and Debye temperatures of MgSiN$_{2}$, MgGeN$_{2}$, MgSnN$_{2}$, AlN, GaN and InN. The experimental references for MgSiN$_{2}$ are evaluated using either polycrystalline elastic constants measured by Bruls {\it et al.}\cite{Bruls2} or measured directly by Bruls {\it et al.}\cite{Bruls2001}. The experimental data for AlN, GaN and InN are evaluated using the single crystal elastic constants presented in Table~\ref{tab:elasticconstants}. The experimental Debye temperatures shown in parenthesis are taken from Wu.\cite{Wu2009}}
\begin{ruledtabular}
\begin{tabular}{lccccc}
 & $\rho$ (g/cm)  & $v_{l}$ (m/s)  & $v_{t}$ (m/s)  & $v_{m}$ (m/s)  & $\Theta_{D}$ (K)\\
 \hline
MgSiN$_{2}$ (PBEsol)& 3.122 & 1.008$\cdot10^4$ & 5.765$\cdot10^3$ & 6.406$\cdot10^3$ & 866 \\
MgSiN$_{2}$ (Expt.\cite{Bruls2}) & 3.143 & 1.032$\cdot10^4$ & 5.996$\cdot10^3$ & 6.653$\cdot10^3$ & 901\\
MgSiN$_{2}$ (Expt.\cite{Bruls2001}) & 3.142 & 1.033$\cdot10^4$ & 5.99$\cdot10^3$ & 6.65$\cdot10^3$ & 900\\
MgGeN$_{2}$ (PBEsol)& 4.411 &  7.828$\cdot10^3$ & 4.394$\cdot10^3$ & 4.890$\cdot10^3$ & 640\\
MgSnN$_{2}$ (PBEsol)& 5.110 & 6.431$\cdot10^3$ & 3.478$\cdot10^3$ & 3.881$\cdot10^3$ & 481 \\
AlN (PBEsol)& 3.258 & 1.056$\cdot10^4$  & 6.066$\cdot10^3$ & 6.737$\cdot10^3$ & 918\\
AlN (Expt.\cite{McNeil}) & 3.258\cite{Wu2009} & 1.092$\cdot10^4$ & 6.396$\cdot10^3$ & 7.092$\cdot10^3$ & 966 (1150\cite{Wu2009})\\
GaN (PBEsol)& 6.131 & 7.375$\cdot10^3$ & 4.205$\cdot10^3$ & 4.673$\cdot10^3$ & 619 \\
GaN (Expt.\cite{Polian}) & 6.089\cite{Wu2009} & 7.799$\cdot10^3$ & 4.444$\cdot10^3$ & 4.939$\cdot10^3$ & 653 (600\cite{Wu2009}) \\
InN (PBEsol)& 6.947 & 5.493$\cdot10^3$ & 2.814$\cdot10^3$ & 3.152$\cdot10^3$ & 377\\
InN (Expt.\cite{Sheleg}) & 6.952\cite{Wu2009} & 4.933$\cdot10^3$ & 1.795$\cdot10^3$ & 2.038$\cdot10^3$ & 244 (660\cite{Wu2009})\\
InN (Expt.\cite{Morales2009}) & 6.952\cite{Wu2009} & 5.671$\cdot10^3$ & 3.001$\cdot10^3$ & 3.354$\cdot10^3$ & 402\\
 \end{tabular}
\end{ruledtabular}
\end{table*}

Based on the knowledge of the the elastic constants it is possible to evaluate the speed of sound of a compound. Here we have calculated the speed of sound of polycrystalline samples as
\begin{equation}
v_{l} = \left[ \frac{B + \frac{4G}{3}}{\rho}\right]^{1/2}
\end{equation}
and
\begin{equation}
v_{t} = \left( \frac{G}{\rho}\right)^{1/2}
\end{equation}
for longitudinal and transverse waves, respectively, and $\rho$ is the mass density of the crystal. The average sound velocity is given by
\begin{equation}
v_{m} = \left[\frac{1}{3}\left(  \frac{2}{v_{t}^3} + \frac{1}{v_{l}^3}  \right) \right]^{1/3}.
\end{equation}
\par
The Debye temperature is an important physical property of a solid. Due to the fact that vibrational excitations arise solely from acoustic vibrations at low temperatures, it is possible to calculate the Debye temperature at low temperatures from elastic constants data as  
\begin{equation}
\Theta_{D} = \frac{h}{k} \left[  \frac{3n}{4\pi} \left(  \frac{N_{A}\cdot\rho}{M}  \right)   \right]^{1/3}v_{m},
\end{equation}
where $h$ and $k$ are Planck's and Boltsmann's constants, respectively, $N_{A}$ is Avogadro's number, $M$ is the Molar mass, and $n$ is the number of atoms in one formula unit, i.e. 4 in the case of MgSiN$_{2}$ and 2 in the case of AlN. In Table~\ref{tab:debye} we present the velocity of sound and the Debye temperatures of the Mg-IV-N$_{2}$ and the III-N systems. We find that the Debye temperature in MgSiN$_{2}$ to be 866~K which is in good agreement with experimental Debye temperatures. In AlN, we find the Debye temperature to be 918~K which is lower than the experimental reference. It is clear, however, that the Debye temperature in MgSiN$_{2}$ is lower than in AlN. As the Mg-IV-N$_{2}$ and III-N systems become heavier the Debye temperatures become significantly reduced.
\par
Regarding the velocity of sound in these compounds, we find that our calculated sound velocities are in very good agreement with available experimental results. We also find that the velocity of sound in MgSiN$_{2}$ is smaller than in AlN, while the sound velocities in MgGeN$_{2}$ and MgSnN$_{2}$ are larger than in GaN and InN respectively. 
\par

\section{Summary and conclusions}
We have performed density functional calculations of the structure and elastic constants of Mg-IV-N$_{2}$ and III-N systems, where the Group IV elements are Si, Ge and Sn while the Group III elements are Al, Ga and In. It is found that MgSiN$_{2}$ should be possible to be grown on both AlN and GaN without significant strain, since the wurtzite-like lattice constant in MgSiN$_{2}$ differs only slightly from the in-plane lattice constants in both AlN and GaN. The same also holds for MgGeN$_{2}$ grown on GaN, however, for such a system the strain will be slightly larger compared to MgSiN$_{2}$ grown on, e.g., AlN. 
\par
The elastic constants in MgSiN$_{2}$ and MgGeN$_{2}$ have been found to be softer than in AlN and GaN, while MgSnN$_{2}$ have a similar hardness to InN. This works towards making it even easier for MgSiN$_{2}$ and MgGeN$_{2}$ to be grown on both AlN and GaN. A small lattice mismatch and softer elastic constants should reduce the number of dislocations and other types of defects if, for example, MgSiN$_{2}$ is grown on GaN compared to AlN grown on GaN.
\par
The elastic anisotropy has also been evaluated and we find that the anisotropy in all investigated nitrides are relatively small and without any large differences between the anisotropy in the Mg-IV-N$_{2}$ and III-N systems.

\section{Acknowledgements}
We acknowledge support from the Leverhulme Trust via M. A. Moram's Research Leadership Award (RL-007-2012). M. A. Moram acknowledges further support from the Royal Society through a University Research Fellowship. This work used the Imperial College high performance computing facilities and, via our membership of the UK's HEC Materials Chemistry Consortium funded by EPSRC (EP/L000202), the ARCHER UK National Supercomputing Service (http://www.archer.ac.uk).

\bibliography{mgsin2}

\end{document}